\documentclass[twocolumn,pra,showpacs]{revtex4}

\usepackage{amssymb} 
\usepackage{amsfonts} 
\usepackage{amsmath} 
\usepackage{graphicx} 
 
\begin{document} 
 
\title{Random Lindblad equations from complex environments} 
\author{Adri\'{a}n A. Budini} 
\affiliation{Max Planck Institute for the Physics of Complex Systems, N{\"o}thnitzer Stra{%
\ss }e 38, 01187 Dresden, Germany} 
\date{\today} 
 
\begin{abstract} 
In this paper we demonstrate that Lindblad equations characterized by a 
random rate variable arise after tracing out a complex structured reservoir. 
Our results follows from a generalization of the Born-Markov approximation, 
which relies in the possibility of splitting the complex environment in a 
direct sum of sub-reservoirs, each one being able to induce by itself a 
Markovian system evolution. Strong non-Markovian effects, which 
microscopically originate from the entanglement with the different 
sub-reservoirs, characterize the average system decay dynamics. As an 
example, we study the anomalous irreversible behavior of a quantum tunneling 
system described in an effective two level approximation. Stretched 
exponential and power law decay behaviors arise from the interplay between 
the dissipative and unitary hopping dynamics. 
\end{abstract} 
 
\maketitle 
 
\section{Introduction} 
 
The dynamic of a small quantum system interacting weakly with uncontrollable 
degrees of freedom is well understood when a Markovian approximation 
applies. In this situation, after tracing-out the environment, the system 
density matrix evolution can be well approximated by a Lindblad equation  
\cite{alicki,nielsen}. 
 
Besides that the applicability of the Markovian approximation range over 
many areas of physics \cite{blum,cohen,carmichael,weiss,lutz} there exist 
several real systems whose dynamics present strong departures from it. The 
main signature of this departure is the presence of strong non-exponential 
decay behaviors, such as power law and stretched exponential. Some examples 
are nanocrystal quantum dots under laser radiation \cite%
{schlegel,brokmann,grigolini}, superconducting qubits \cite%
{makhlinReport,makhlin,falci}, spin environments \cite{dobro}, dephasing in 
atomic and molecular physics \cite{gruebele}, electron transfer and exciton 
dynamics in proteins \cite{klein}, and molecular systems maintained in a 
glassy environment \cite{wong}, to name but a few. These and another 
specific experimental situations rise up the necessity of finding formalism 
and effective evolutions able to describe the corresponding non-Markovian 
dynamics. 
 
When the environment is modeled as an infinite set of normal modes, 
departure from a Markov approximation can be related to the corresponding 
spectral density function. This situation was extensively studied for the 
spin-boson and boson-boson models, where exacts solutions are available \cite%
{leggett,grifoni,lang,hanggi,milena}. Exact solutions can also be formulated 
for more general systems. Nevertheless, due to the huge analytical and 
numerical efforts needed for getting the non-Markovian system dynamics, 
alternative numerical methods based in a special decomposition of the 
spectral density function were formulated \cite{tannor,ulrich}. 
 
Anomalous system dynamics also arise from a random matrix modeling of the 
system-environment interaction \cite{kuznezov,DCohen,gaspard,kottos}. This 
approach naturally applies when describing environments characterized by a 
complex dynamics as for example chaotic ones. 
 
Outside from a microscopic point of view, there exist an increasing interest 
in describing non-Markovian effects in open quantum systems by introducing 
memory contributions in Lindblad evolutions \cite%
{wilkie,budini,cresser,lidar}. This procedure provide easy manageable 
dynamics. While most of these models are phenomenological, in this paper we 
will relate the presence of strong memory effects in the standard Lindblad 
theory \cite{budini}\ with the microscopic interaction of a system with a 
complex structured environment. 
 
We will base our considerations in a splitting of the full Hilbert space of 
the bath in a direct sum of sub-reservoirs, constructed in such a way that 
each one guarantees the conditions for the applicability of a Markov 
approximation. Our motivation for formulating this splitting comes from 
systems embedded in glassy environments, where the underlying disorder 
produce localized bath states, inducing a natural shell structure of modes, 
each set having a different coupling strength with the system \cite{wong}. 
Thus, we associate a different Markovian sub-bath to each set of states. 
 
As we will demonstrate, the splitting assumption allows us to generalize the 
usual Born-Markov approximation, which in a natural way leads to the 
formulation of Lindblad equations characterized by a random dissipative 
rate. As is well known from a classical context, master equations with 
random rates \cite{klafter,bernasconi,odagaki,miguel} are useful for 
describing strong non-Markovian effects \cite{metzler,CoolingBook}. Here we 
will demonstrate that the same scheme can also be applied in a quantum 
context under the previous conditions, i.e., a complex environment under the 
splitting condition. 
 
As an example, we will study the anomalous dissipative dynamics of a quantum 
tunneling system described in a two level approximation \cite{leggett,hanggi}%
. Strong non-exponential behaviors, such as stretched exponential and power 
law, arise from the interplay between the unitary hopping dynamics and the 
memory effects induced by the environment. The conditions under which our 
modeling can be mapped to the spin-boson model and stochastic dynamics are 
established. In this context, the differences between our framework and 
other approaches \cite{tannor,ulrich} introduced to deal with non-Markovian 
environments are established. 
 
\section{Reduced dynamics from complex environments} 
 
In general, the evolution of a system interacting with a complex environment 
can not be described in a Markovian approximation. While a general treatment 
is clearly not possible, when the splitting condition applies, for weak 
system-bath interactions, the reduced dynamics can be described through a 
generalization of the Born-Markov approximation. 
 
\subsection{Generalized Born-Markov approximation} 
 
We start by assuming a full microscopic Hamiltonian description%
\begin{equation} 
H_{T}=H_{S}+H_{B}+H_{I}, 
\end{equation}%
where $H_{S}$ correspond to the Hamiltonian of a system S, and $H_{B}$ 
correspond to the Hamiltonian of the bath B. The term $H_{I}=q_{S}\otimes 
Q_{B}$\ describes their mutual interaction, with the operators $q_{S}$ and $%
Q_{B}$ acting on the system and bath Hilbert spaces respectively. 
 
In an interaction representation with respect to $H_{S}+H_{B}$, the total 
density matrix $\rho _{T}(t)$ evolves as%
\begin{equation} 
\frac{d\rho _{T}(t)}{dt}=\frac{-i}{\hbar }[H_{I}(t),\rho _{T}(t)], 
\end{equation}%
where $H_{I}(t)$ is the interaction Hamiltonian in the Heisenberg 
representation. Integrating formally this equation, and substituting the 
solution in it, the evolution of the reduced system density matrix $\rho 
_{S}(t)=\mathrm{Tr}_{B}\{\rho _{T}(t)\}$ can be written as%
\begin{equation} 
\frac{d\rho _{S}(t)}{dt}=-\left( \frac{1}{\hbar }\right) 
^{2}\int_{0}^{t}dt^{\prime }\mathrm{Tr}_{B}\{[H_{I}(t),[H_{I}(t^{\prime 
}),\rho _{T}(t^{\prime })]]\},  \label{Exact} 
\end{equation}%
where, as usual, a first order contribution was discarded after assuming $%
\mathrm{Tr}_{B}[H_{I}(t)\rho _{T}(0)]=0$. From this evolution, the well 
known Born-Markov approximation can be deduced \cite%
{blum,cohen,carmichael,weiss}. The Born approximation consists into assume, 
at all times, an uncorrelated structure for the total density matrix%
\begin{equation} 
\rho _{T}(t)\approx \rho _{S}(t)\otimes \rho _{B},  \label{Born} 
\end{equation}%
where $\rho _{B}$ define a stationary state of the bath. This assumption is 
consistent up to second order in the interaction Hamiltonian. When the decay 
of the bath correlation defines the small time-scale of the problem, after 
introducing Eq.~(\ref{Born}) in Eq.~(\ref{Exact}), the Markovian 
approximation leads to a closed local in time density matrix evolution. 
 
We remark that the Born-Markov approximation does not rely in any specific 
model of bath dynamics \cite{blum,cohen,carmichael,weiss}, such as an 
infinite set of harmonic oscillators. In fact, its master equation is 
independent of model assumptions used in its derivation \cite{lutz}. 
 
Now we consider a complex environment for which the previous approximations 
are not valid. As is usual when dealing with complex environments \cite%
{kuznezov,DCohen,gaspard,kottos}, instead of defining the bath Hamiltonian $%
H_{B}$ as an infinite set of normal modes, here we specify it through its 
eigenstates basis $\{\left\vert \epsilon \right\rangle \}$, which in a weak 
interaction limit, results unmodified by the interaction with the system. As 
a central hypothesis, we will assume that, while the full action of the 
environment can not be described in a Markov approximation, it is possible 
to split the full Hilbert space of the bath as a direct sum of subspaces, in 
such a way that each one defines a sub-reservoir able to induce by itself a 
Markovian system evolution \cite{footnotebath}. These hypotheses are the 
main assumptions that allow us to formulate our results. 
 
In conformity with the splitting condition, we write the interaction 
Hamiltonian as a direct sum of sub-Hamiltonians%
\begin{equation} 
H_{I}=H_{I_{1}}\oplus H_{I_{2}}\cdots \oplus H_{I_{R}}\oplus 
H_{I_{R+1}}\cdots , 
\end{equation}%
where $H_{I_{R}}=q_{S}\otimes Q_{B_{R}}.$ Here, each operator $Q_{B_{R}}$ 
defines the interaction between the system and each sub-reservoir $R.$ 
 
In order to describe the joint action of all sub-reservoirs over the system, 
instead of the uncorrelated form Eq.~(\ref{Born}), we introduce the \textit{%
generalized Born-approximation}%
\begin{equation} 
\rho _{T}(t)\approx \sum_{R}\rho _{R}(t)\otimes \Xi _{R},  \label{RhoTot} 
\end{equation}%
where $\mathrm{Tr}_{S}[\rho _{R}(t)]=1,$ and we have defined%
\begin{equation} 
\Xi _{R}=\sum_{\{\epsilon _{R}\}}\left\langle \epsilon _{R}\right\vert \rho 
_{B}\left\vert \epsilon _{R}\right\rangle \left\vert \epsilon 
_{R}\right\rangle \left\langle \epsilon _{R}\right\vert . 
\end{equation}%
$\{\left\vert \epsilon _{R}\right\rangle \}$ is the base of eigenvectors 
that span the subspace corresponding to each sub-reservoir. Therefore, each 
contribution in Eq.~(\ref{RhoTot}) consists in the external product between 
a system state $\rho _{R}(t)$ and the projection of the stationary bath 
state $\rho _{B}$ over each subspace $R$. In physical terms, each state $%
\rho _{R}(t)$ takes in account the dissipative effects induced by each 
sub-reservoir. 
 
After introducing Eq.~(\ref{RhoTot}) in Eq.~(\ref{Exact}), we get the 
approximated evolution%
\begin{eqnarray} 
\frac{d\rho _{S}(t)}{dt} &\approx &-\left( \frac{1}{\hbar }\right) 
^{2}\sum_{R}P_{R}\int_{0}^{t}dt^{\prime }  \label{BornGeneral} \\ 
&&\mathrm{Tr}_{B_{R}}\{[H_{I_{R}}(t),[H_{I_{R}}(t^{\prime }),\rho 
_{R}(t^{\prime })\otimes \rho _{B_{R}}]]\},  \notag 
\end{eqnarray}%
where $\mathrm{Tr}_{B_{R}}\{\bullet \}$ means a trace operation with the 
states $\{\left\vert \epsilon _{R}\right\rangle \}$ corresponding to each 
subspace. Furthermore, we have introduced the sub-bath density states $\rho 
_{B_{R}}=\Xi _{R}/P_{R},$ where%
\begin{equation} 
P_{R}=\mathrm{Tr}_{B_{R}}\{\Xi _{R}\}=\sum_{\{\epsilon _{R}\}}\left\langle 
\epsilon _{R}\right\vert \rho _{B}\left\vert \epsilon _{R}\right\rangle . 
\label{pesos} 
\end{equation}%
The normalization condition $\mathrm{Tr}_{B}[\rho _{B}]=1$ implies the 
relation $\sum_{R}P_{R}=1$. Thus, the set $\{P_{R}\}$ can be seen as a set 
of probabilities defined by the weight of each subspace in the total 
stationary bath state. 
 
From Eq.~(\ref{RhoTot}) we can write%
\begin{equation} 
\rho _{S}(t)=\mathrm{Tr}_{B}[\rho _{T}(t)]\approx \sum_{R}P_{R}\rho _{R}(t). 
\label{sum} 
\end{equation}%
Then, the evolution Eq.~(\ref{BornGeneral}) is in fact a linear combination 
of the evolutions corresponding to the set $\{\rho _{R}(t)\}$, each one 
participating with weight $P_{R}$. After introducing the \textit{Markovian 
approximation} \cite{blum,cohen,carmichael,weiss} to the evolution of each 
state $\rho _{R}(t)$,\ in a Schr\"{o}dinger representation, we get%
\begin{eqnarray} 
\frac{d\rho _{R}(t)}{dt} &=&\frac{-i}{\hbar }[H_{S},\rho _{R}(t)]-\left(  
\frac{1}{\hbar }\right) ^{2}\int_{0}^{\infty }dt^{\prime }  \label{markov} \\ 
&&\mathrm{Tr}_{B_{R}}\{[H_{I_{R}},[H_{I_{R}}(-t^{\prime }),\rho 
_{R}(t)\otimes \rho _{B_{R}}]].  \notag 
\end{eqnarray}%
This evolution correspond to the usual Born-Markov approximation when 
considering a bath consisting only of the subset of states $\{\left\vert 
\epsilon _{R}\right\rangle \}$ and characterized by the stationary state $%
\rho _{B_{R}}$. The system density matrix is defined by the linear 
combination Eq.~(\ref{sum}). 
 
\subsection{Random Lindblad equations} 
 
The evolution Eq.~(\ref{markov}), disregarding transients of the order of 
the sub-bath Hamiltonian correlation time, can be always well approximated 
by a Lindblad equation \cite{alicki}%
\begin{equation} 
\frac{d\rho _{R}(t)}{dt}=\mathcal{L}_{H}[\rho _{R}(t)]+\gamma _{R}\mathcal{L}%
[\rho _{R}(t)],  \label{Lindblad} 
\end{equation}%
where $\mathcal{L}_{H}[\bullet ]=(-i/\hbar )[H_{S},\bullet ]$ is the system 
Liouville superoperator and the Lindblad superoperator is defined by%
\begin{equation} 
\mathcal{L}[\bullet ]=\sum_{\alpha }\frac{1}{2}([V_{\alpha },\bullet 
V_{\alpha }^{\dagger }]+[V_{\alpha }\bullet ,V_{\alpha }^{\dagger }]). 
\end{equation}%
As the underlying microscopic interaction between the system and the 
environment is the same in each subspace, the set of operators $\{V_{\alpha 
}\}$\ does not depend on index $R$. Nevertheless, each subspace has 
associated a different characteristic dissipative rate $\gamma _{R}$. As 
this rate arises from the interaction of the system with the manifold of 
states $\{\left\vert \epsilon _{R}\right\rangle \}$, consistently with the 
Fermi golden rule \cite{cohen}, it is proportional to the characteristic 
interaction strength of each subspace, denoted as $|Q_{B_{R}}|,$ multiplied 
by the corresponding density of states $g_{R}(\epsilon )=\sum_{\{\epsilon 
_{R}\}}\delta (\epsilon -\epsilon _{R})$ evaluated in a characteristic 
frequency $\omega _{S}$ of the system, i.e., $\gamma _{R}\approx 
|Q_{B_{R}}|^{2}g_{R}(\hbar \omega _{S})$. 
 
With these definitions in hand, we conclude that under the generalized 
Born-Markov approximation, we can represent the dynamics induced by the 
complex environment by\ a Lindblad master equation characterized by a random 
rate variable, defined by the set $\{\gamma _{R},P_{R}\}$. Correspondingly, 
the system state follows from the average%
\begin{equation} 
\rho _{S}(t)=\sum_{R}P_{R}\rho _{R}(t)\equiv \langle \rho _{R}(t)\rangle . 
\label{average} 
\end{equation}%
Random rate equations were extensively used to model classical anomalous 
diffusion processes in disordered media \cite%
{klafter,bernasconi,odagaki,miguel}. Here, we have derived a similar 
structure for a different physical situation, i.e., quantum systems embedded 
in a complex structured environment. 
 
As in a classical context, while the density matrixes $\rho _{R}(t)$ follows 
a Markovian evolution, the average state $\rho _{S}(t)$ evolve with a 
non-Markovian evolution. This evolution can be easily obtained in a Laplace 
domain, where the average Eq.~(\ref{average}) takes the form%
\begin{equation} 
\rho _{S}(u)=\left\langle \frac{1}{u-(\mathcal{L}_{H}+\gamma _{R}\mathcal{L})%
}\right\rangle \rho _{S}(0)\equiv \left\langle G_{R}(u)\right\rangle \rho 
_{S}(0),  \label{rholaplace} 
\end{equation}%
with $u$ being the Laplace variable, and we have used the solutions $\rho 
_{R}(t)=\exp [(\mathcal{L}_{H}+\gamma _{R}\mathcal{L})t]\rho _{S}(0).$ 
Consistently with the uncorrelated initial condition $\rho _{T}(0)=\rho 
_{S}(0)\otimes \rho _{B}(0)$, there not exist any statistical correlation 
between $\rho _{S}(0)$ and the random rate set. Thus, the average evolution 
can be obtained without appealing to a projector technique \cite%
{klafter,bernasconi,odagaki,miguel}. In fact, after introducing in Eq.~(\ref%
{rholaplace}) the identity in the form $\langle G_{R}(u)[u-(\mathcal{L}%
_{H}+\gamma _{R}\mathcal{L})]\rangle ^{-1}$, it is immediate to get the 
deterministic, closed, non-Markovian evolution equation%
\begin{equation} 
\frac{d\rho _{S}(t)}{dt}=\mathcal{L}_{H}[\rho _{S}(t)]+\int_{0}^{t}d\tau \,%
\mathbb{L}(t-\tau )[\rho _{S}(\tau )],  \label{noMarkov} 
\end{equation}%
where the superoperator $\mathbb{L}$ is defined in the Laplace domain by the 
equation%
\begin{equation} 
\langle G_{R}(u)\gamma _{R}\mathcal{L}\rangle \lbrack \bullet ]=\langle 
G_{R}(u)\rangle \mathbb{L}(u)[\bullet ].  \label{memory} 
\end{equation}%
Depending on the set $\{\gamma _{R},P_{R}\}$, Eq.~(\ref{noMarkov}) may lead 
to the presence of strong non-Markovian decay behaviors in the system 
dynamics. This characteristic originates from the entanglement of the system 
which each sub-reservoir, situation explicitly introduced by Eq.~(\ref%
{RhoTot}). 
 
An example of complex structured environment where the generalized 
Born-Markov applies straightforwardly is a bath Hamiltonian whose 
eigenvectors can be labelled with two indexes $(E,R)$. The index $E$ is 
continuous, and for each $R$ the corresponding sub-manifold of states is 
able to induce a different system Markovian-decay. The difference between 
the Markovian dynamics may originates in the coupling strength of each 
sub-manifold with the system. On the other hand, it may originates due to 
strong variations of the bath density of states with index $R$. The system 
dynamics follows as a superposition of Markovian dynamics whose weights are 
taken in account through the generalized Born approximation Eq.~(\ref{RhoTot}%
). If the decay induced by each sub-manifold is the same, the generalized 
Born approximation reduces to the usual one, and then a Markovian evolution 
is obtained. Further examples can be established in the context of random 
band-matrix models \cite{lutz}, where the Markovian sub-baths, for example, 
may be associated to subspaces with a different characteristic bandwidth. 
 
\subsection{Effective approximation} 
 
Classical master equations with random rates are characterized by equations 
similar to those obtained previously. Nevertheless, as in general the 
underlying numbers of states is infinite, some kind of approximation is 
necessary in order to obtain the operator $\mathbb{L}$, as for example an 
effective medium approximation \cite{odagaki,miguel}. Here, we introduce a 
similar approximation in order get a general characterization of the 
dynamics. Thus, in Eq.~(\ref{memory}) we discard the dependence introduced 
by the Lindblad superoperator $\mathcal{L}$ in the propagator $G_{R}(u)$, 
i.e., $\mathcal{L}\rightarrow -$\textrm{I}. Then, we get the approximated 
solution $\mathbb{L}(u)\simeq K(u-\mathcal{L}_{H})\mathcal{L},$ from where 
it follows the evolution%
\begin{equation} 
\frac{d\rho _{S}(t)}{dt}\simeq \mathcal{L}_{H}[\rho 
_{S}(t)]+\int_{0}^{t}d\tau K(t-\tau )e^{(t-\tau )\mathcal{L}_{H}}\mathcal{L}%
[\rho _{S}(\tau )].  \label{aproximada} 
\end{equation}%
In this approximation all information about the random rate is introduced 
through the kernel function%
\begin{equation} 
K(u)=\left\langle \frac{\gamma _{R}}{u+\gamma _{R}}\right\rangle 
\left\langle \frac{1}{u+\gamma _{R}}\right\rangle ^{-1}.  \label{K(u)} 
\end{equation}%
As in a classical context, this kernel can be associated with a waiting time 
distribution $w(t)$\ and a survival probability $P_{0}(t)$\ defined by%
\begin{equation} 
w(u)=\left\langle \frac{\gamma _{R}}{u+\gamma _{R}}\right\rangle ,\ \ \ \ \ 
P_{0}(u)=\left\langle \frac{1}{u+\gamma _{R}}\right\rangle . 
\label{survival} 
\end{equation}%
In classical master equations, these objects define a continuous time random 
walk \cite{klafter,bernasconi,odagaki,miguel,metzler,CoolingBook}. In the 
quantum case, a similar stochastic dynamics can be constructed \cite{budini}%
. It consists in the application at random times of the superoperator $%
\mathcal{E}=\mathcal{L}+$\textrm{I}, implying the transformation $\rho 
\rightarrow \mathcal{E}[\rho ]$, while during the intervals between these 
disruptive actions the system evolves with its unitary dynamics, $U(t)=\exp 
[t\mathcal{L}_{H}]$. The intervals between the successive applications of $%
\mathcal{E}$ follows from the waiting time distribution $w(t)$. The function  
$\ P_{0}(t)$ defines the corresponding survival probability, $%
P_{0}(t)=1-\int_{0}^{t}d\tau w(\tau )$. Thus, the average over different 
realizations of the random times can be written as%
\begin{eqnarray} 
\rho _{S}(t) &=&P_{0}(t)e^{t\mathcal{L}_{H}}\rho _{S}(0)  \notag \\ 
&&+\int_{0}^{t}d\tau \,w(t-\tau )e^{(t-\tau )\mathcal{L}_{H}}\mathcal{E}%
[\rho _{S}(\tau )].  \label{estocastica} 
\end{eqnarray}%
From here, in a Laplace domain, it is straightforward to recuperate the 
evolution Eq.~(\ref{aproximada}). When $\mathcal{L}\neq \mathcal{E}-$I, with  
$\mathcal{E}$ a completely positive superoperator \cite{nielsen}, a similar 
stochastic dynamics can be formulated after introducing a limit procedure  
\cite{footnote0}. 
 
We remark that the stochastic interpretation [Eq.~(\ref{estocastica})] was 
constructed after associating to the kernel $K(t)$ a waiting time 
distribution and a survival probability, Eq.~(\ref{survival}). This 
association does not rely in the generalized Born-Markov approximation, 
neither it was deduced from a conditional continuous time measurement theory~%
\cite{carmichael}. Therefore, it is not clear if one can associate to the 
stochastic dynamics a random signal of a measurement apparatus. If this is 
the case, contradictions between environmental decoherence and wave-function 
collapse may arise~\cite{bologna,pala}. 
 
\section{Quantum tunneling system driven by a complex environment} 
 
As an example of our formalism, in this section we will characterize the 
dissipative dynamics of a quantum tunneling system described in a two level 
approximation \cite{leggett,hanggi} and driven by a complex environment. 
Then, the system Hamiltonian can be written as  
\begin{equation} 
H_{S}=\frac{\hbar \omega _{A}}{2}\sigma _{z}+\frac{\hbar \Delta }{2}\sigma 
_{x}. 
\end{equation} 
The first term, proportional to the $z$-Pauli matrix $\sigma _{z}$ define 
the energy of the effective levels, and the second one, proportional to the $%
x$-Pauli matrix $\sigma _{x}$, introduce the reversible hopping between the 
two effective states. 
 
The complex environment will be represented by the Lindblad superoperator%
\begin{equation} 
\mathcal{L}[\bullet ]=\frac{1}{2}([\sigma _{z}\bullet ,\sigma _{z}]+[\sigma 
_{z},\bullet \sigma _{z}]),  \label{lindblad} 
\end{equation}%
and an arbitrary set ${\{\gamma _{R},P_{R}\}}$ of random rates and weights. 
For fixed rate, this superoperator induces a dynamics equivalent to a 
thermal environment in a high temperature limit \cite{milena}. 
 
The evolution of the system density matrix is defined by Eqs.~(\ref{noMarkov}%
) and (\ref{memory}). Here, we write the evolution in terms of the 
components of the Bloch vector, which are defined by the mean value of the 
Pauli matrixes, $S_{j}(t)=\mathrm{Tr}_{S}\{\rho _{S}(t)\sigma _{j}\}$, with $%
j=x,$ $y,$ and $z$. We get  
\begin{subequations} 
\label{bloch} 
\begin{eqnarray} 
\frac{dS_{X}(t)}{dt} &=&-\omega _{A}S_{Y}(t)-\int_{0}^{t}d\tau \{\Gamma 
_{X}(t-\tau )S_{X}(\tau )\ \ \ \ \ \  \\ 
&&-\Upsilon (t-\tau )S_{Y}(\tau )\},  \notag \\ 
\frac{dS_{Y}(t)}{dt} &=&\omega _{A}S_{X}(t)-\Delta S_{Z}(t)-\int_{0}^{t}d\tau 
\\ 
&&\{\Gamma _{Y}(t-\tau )S_{Y}(\tau )+\Upsilon (t-\tau )S_{X}(\tau )\},  
\notag \\ 
\frac{dS_{Z}(t)}{dt} &=&\Delta S_{Y}(t). 
\end{eqnarray}%
Thus, the system evolution is completely characterized by three memory 
kernels $\Gamma _{X}(t)$, $\Gamma _{Y}(t)$, and $\Upsilon (t)$. In Appendix 
A, we give the exact expressions of these kernels for arbitrary random 
rates, joint with the kernels that arise from the effective approximation 
Eq.~(\ref{aproximada}). From Eq.~(\ref{bloch}) it is straightforward to 
write the system density matrix evolution [Eq.~(\ref{noMarkov})] as a sum of 
Lindblad superoperators, each one characterized by a different kernel. 
 
\subsection{Dispersive limit} 
 
When the hopping frequency is zero, $\Delta =0$, the dynamics reduce to a 
dispersive one. Thus, the coherences decay continuously while the population 
of each effective level remains constant. In this limit, from Appendix A, 
for arbitrary set $\{\gamma _{R},P_{R}\}$ we get the exact kernels  
\end{subequations} 
\begin{subequations} 
\label{kernelDispersivo} 
\begin{eqnarray} 
\Gamma _{X}(t) &=&K(t)\cos [\omega _{A}t], \\ 
\Gamma _{Y}(t) &=&K(t)\cos [\omega _{A}t], \\ 
\Upsilon (t) &=&K(t)\sin [\omega _{A}t]. 
\end{eqnarray}%
where $K(t)$ is defined in the Laplace domain by Eq.~(\ref{K(u)}). We note 
that these kernels also arise from the effective approximation Eq.~(\ref%
{aproximada}), indicating that for $\Delta =0$, both evolutions coincide. 
 
From Eqs.~(\ref{bloch}), the exact solution of the Bloch vector is given by  
\end{subequations} 
\begin{subequations} 
\label{disperso} 
\begin{eqnarray} 
S_{X}(t) &=&P_{0}(t)\{\cos [\omega _{A}t]S_{X}(0)-\sin [\omega 
_{A}t]S_{Y}(0)\},\ \ \ \ \ \ \ \  \\ 
S_{Y}(t) &=&P_{0}(t)\{\sin [\omega _{A}t]S_{X}(0)+\cos [\omega 
_{A}t]S_{Y}(0)\},\ \ \ \ \ \ \ \  \\ 
S_{Z}(t) &=&S_{Z}(0), 
\end{eqnarray}%
where $P_{0}(t)$ is the survival probability defined by its Laplace 
transform Eq.~(\ref{survival}), which in the time domain reads $%
P_{0}(t)=\sum_{R}P_{R}\exp [-\gamma _{R}t].$ Consistently, we note that the 
exact solutions Eqs.~(\ref{disperso}) correspond to an average over 
Markovian solutions, each one characterized by a rate $\gamma _{R}$ and 
participating with weight $P_{R}$. Depending on the distribution of the 
dissipation rate, arbitrary forms of the decay can be obtained from this 
average over exponential functions. Hence the non-Markovian behavior can be 
observed in the relaxation of the density matrix to the stationary state. 
 
\subsection{Anomalous decay behaviors} 
 
The form of the set $\{\gamma _{R},P_{R}\}$ depends on the specific 
structure of the complex environment. Here, we will determine this set in a 
phenomenological way as a function of the system decay behavior. We will be 
interested in obtaining anomalous decay dynamics such as \textit{power law}. 
A possible set consistent with this decay is  
\end{subequations} 
\begin{equation} 
\gamma _{R}=\gamma _{0}\exp [-bR],\;\;\;\;\;\;P_{R}=(1-e^{-a})\exp [-aR], 
\label{expo} 
\end{equation}%
where $R\in \lbrack 0,\infty ],$ $\gamma _{0}$ scale the random rates, and 
the constants $b$ and $a$ measure the exponential decay of the random rates 
and their corresponding weights. With these definitions, it is simple to 
demonstrate that after a transient of order $\gamma _{0}$, the waiting time 
distribution and its associated survival probability, Eq.~(\ref{survival}), 
present a power law decay behavior \cite{alemany}, $w(t)\approx 1/(\gamma 
_{0}t)^{1+\alpha }$, and $P_{0}(t)\approx 1/(\gamma _{0}t)^{\alpha },$ where  
$\alpha =a/b$. Clearly, this behavior is reflected in the system dynamics. 
 
\begin{figure}[tbph] 
\includegraphics[height=7.2cm]{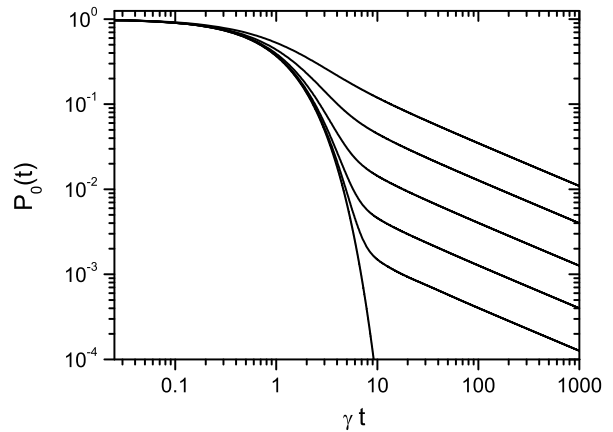} 
\caption{Survival probability. From top to bottom, the parameters are $%
\protect\beta /\protect\gamma =0.75$, $10^{-1}$, $10^{-2}$, $10^{-3}$, $%
10^{-4}$, and the Markovian limit $\protect\beta =0$. In all cases we take $%
\protect\alpha =1/2$.} 
\label{Figura1} 
\end{figure} 
When $0<\alpha <1$, the kernel $K(t)$ corresponding to the set Eq.~(\ref%
{expo}) can be well approximated by the expression%
\begin{equation} 
K(u)\simeq \frac{\gamma }{1+(\beta /u)^{1-\alpha }},  \label{kernel} 
\end{equation}%
with the definitions%
\begin{equation} 
\gamma =\langle \gamma _{R}\rangle ,\ \ \ \ \ \ \ \ \beta =\frac{\langle 
\gamma _{R}^{2}\rangle -\langle \gamma _{R}\rangle ^{2}}{\langle \gamma 
_{R}\rangle }. 
\end{equation} 
The scaling of these parameters can be motivated by considering a two 
dimensional set of random rates \cite{footnote1}. From Eq.~(\ref{K(u)}) and (%
\ref{survival}), the waiting time distribution and its associated survival 
probability can be obtained as  
\begin{equation} 
w(u)=\frac{K(u)}{u+K(u)},\ \ \ \ \ \ \ \ \ P_{0}(u)=\frac{1}{u+K(u)}. 
\label{sur} 
\end{equation}%
From here, is it is simple to proof that $w(u)$ is a completely monotone 
function \cite{budini}, which implies that $P_{0}(t)$ decays in a monotonous 
way or equivalently, $w(t)\geq 0$. 
 
In Figure 1 we plot the survival probability $P_{0}(t)$ by assuming the 
kernel Eq.~(\ref{kernel}) for different values of $\beta /\gamma $. We note 
that in a short time regime, the decay is an exponential one, while in an 
asymptotic regime a power law behavior is present%
\begin{equation} 
P_{0}(t)\simeq \exp [-\gamma t],\ \ \ \ \ \ \ \ \ \ P_{0}(t)\simeq \frac{%
\beta ^{1-\alpha }}{\gamma \Gamma (1-\alpha )}\frac{1}{t^{\alpha }}, 
\end{equation}%
where $\Gamma (x)$ is the gamma function. These asymptotic behaviors follows 
immediately from Eq.~(\ref{sur}). When the dispersion of the random rate $%
\gamma _{R}$ is zero $(\beta =0)$, consistently the dynamics reduce to a 
Markovian one, $K(u)=\gamma $, which implies the pure exponential decay $%
P_{0}(t)=\exp [-\gamma t]$ and $w(t)=\gamma \exp [-\gamma t]$. 
 
In the next subsection we will characterize the tunneling dynamics by 
assuming a complex environment characterized by the random rate set Eq.~(\ref%
{expo}) or equivalently by the kernel Eq.~(\ref{kernel}). 
 
\subsection{Tunneling dynamics} 
 
Here we will analyze the tunneling dynamics for a symmetric case $\omega 
_{A}=0$, which arise when the two effective levels have the same energy. 
From Appendix A, the exact kernels read  
\begin{subequations} 
\label{kernelHoping} 
\begin{eqnarray} 
\Gamma _{X}(u) &=&K(u), \\ 
\Gamma _{Y}(u) &=&K(u+\Delta ^{2}/u), \\ 
\Upsilon (u) &=&0. 
\end{eqnarray}%
As before, the kernel $K(u)$ is defined by Eq.~(\ref{K(u)}). The exact 
solution of the Bloch vector can be obtained in a Laplace domain. We get  
\end{subequations} 
\begin{subequations} 
\label{hoping} 
\begin{eqnarray} 
S_{X}(u) &=&\frac{1}{u+\Gamma _{X}(u)}S_{X}(0), \\ 
S_{Y}(u) &=&\Lambda (u)\{uS_{Y}(0)-\Delta S_{Z}(0)\}, \\ 
S_{Z}(u) &=&\Lambda (u)\{[u+\Gamma _{Y}(u)]S_{Z}(0)+\Delta S_{Y}(0)\},\ \ \ 
\ \ \  
\end{eqnarray}%
where we have defined  
\end{subequations} 
\begin{equation} 
\Lambda (u)=\frac{1}{u^{2}+u\Gamma _{Y}(u)+\Delta ^{2}}, 
\end{equation}%
which can also be expressed as $\Lambda (u)=P_{0}(u+\Delta ^{2}/u)/u$. 
 
In this case it is not possible to find in the time domain a general exact 
solution for arbitrary memory kernels. A simple analytical solution is only 
available in a Markovian case [$K(u)=\gamma $]  
\begin{subequations} 
\label{Markovian} 
\begin{eqnarray} 
S_{X}(t) &=&e^{-\gamma t}S_{X}(0), \\ 
S_{Y}(t) &=&e^{-\gamma t/2}\{S_{Y}(0)\cosh [\lambda t]  \notag \\ 
&&-\lambda ^{-1}[(\gamma /2)S_{Y}(0)+\Delta S_{Z}(0)]\sinh [\lambda t],\ \ \ 
\ \  \\ 
S_{Z}(t) &=&e^{-\gamma t/2}\{S_{Z}(0)\cosh [\lambda t]  \notag \\ 
&&+\lambda ^{-1}[(\gamma /2)S_{Z}(0)+\Delta S_{Y}(0)]\sinh [\lambda t],\ \ \ 
\ \  
\end{eqnarray}%
where $\lambda =\sqrt{(\gamma /2)^{2}-\Delta ^{2}}$, and $\gamma $ defines 
the unique dissipative rate. Notice that in the limit of null dissipation, a 
periodic hopping between the effective levels is obtained.  
\begin{figure}[tbph] 
\includegraphics[height=6.8cm]{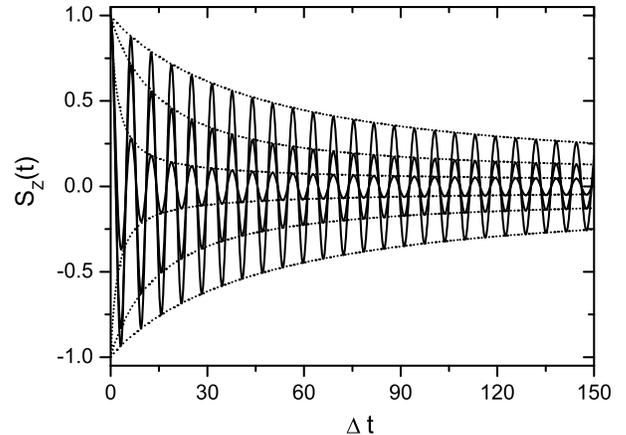} 
\caption{Average $S_{Z}(t)$ considering as initial condition the upper 
eigenstate of $\protect\sigma _{z}$. The envelopes are given by $\pm 
P_{0}(t/2)$. From top to bottom, the parameters are $\protect\gamma /\Delta 
=0.05$, $0.15$, and $1.0$. In all cases we take $\protect\alpha =1/2,$ $%
\protect\beta =\protect\gamma /2$, and $\protect\omega _{A}=0.$} 
\end{figure} 
 
For arbitrary random rates $\{\gamma _{R},P_{R}\}$, the dynamics can be 
characterized in different regimes. First, in the case $\Delta \gg \langle 
\gamma _{R}\rangle $, from Eqs.~(\ref{hoping}) it is possible to get the 
approximated solutions  
\end{subequations} 
\begin{subequations} 
\label{aproxHoping} 
\begin{eqnarray} 
S_{X}(t) &=&P_{0}(t)S_{X}(0),\ \ \ \ \  \\ 
S_{Y}(t) &\simeq &P_{0}(t/2)\{\cos [\Delta t]S_{Y}(0)-\sin [\Delta 
t]S_{Z}(0)\},\ \ \ \ \ \ \  \\ 
S_{Z}(t) &\simeq &P_{0}(t/2)\{\sin [\Delta t]S_{Y}(0)+\cos [\Delta 
t]S_{Z}(0)\}.\ \ \ \ \ \ \  
\end{eqnarray}%
Thus, the dynamics consist in a periodic tunneling between the two effective 
states, and whose decay can be written in terms of the survival probability. 
As in the previous case, this solution correspond to an average over the 
corresponding Markovian solutions, i.e., Eq.~(\ref{Markovian}) written in 
the limit of small decay rate when compared to the tunneling frequency $%
\Delta $. 
 
In Figure 2 we plot the average of the $z$-Pauli matrix which follows from 
Eq.~(\ref{hoping}) with the kernel Eq.~(\ref{kernel}). As initial condition 
we take the upper eigenstate of $\sigma _{z}$. We verified that the exact 
solutions are well described by the approximation Eq.~(\ref{aproxHoping}) 
for parameters values satisfying $\gamma /\Delta \lesssim 1$. As the 
envelope decay is given by $P_{0}(t/2)$, by increasing the average rate $%
\gamma $, the dynamics decay in a faster way. This dependence is broken when 
the average rate is much greater than the hopping frequency. 
 
In the limit $\Delta \ll \langle \gamma _{R}\rangle $, the dissipative 
dynamics dominates over the tunneling one. In Figure 3 we plot $S_{Z}(t)$ 
[Eq.~(\ref{hoping})] for different values of the characteristic parameters 
of the kernel Eq.~(\ref{kernel}). We note that by increasing the average 
rate $\gamma $, a slower decay is obtained. Thus, the dynamics develops a 
Zeno-like effect \cite{zeno,schulman}. From the exact solution Eq.~(\ref%
{hoping}), the characteristic decay of the Bloch vector can be approximated 
by the expressions  
\end{subequations} 
\begin{subequations} 
\begin{eqnarray} 
S_{X}(u) &=&P_{0}(u)S_{X}(0), \\ 
S_{Y}(u) &\simeq &\mathcal{Z}(u)\{uS_{Y}(0)-\Delta S_{Z}(0)\}d(u),\ \ \  \\ 
S_{Z}(u) &\simeq &\mathcal{Z}(u)\{S_{Z}(0)+\Delta S_{Y}(0)d(u)\}, 
\end{eqnarray}%
where we have introduced $\mathcal{Z}(u)=u^{-1}w(\Delta ^{2}/u)$ and the 
function $d(u)=[u+K(\Delta ^{2}/u)]^{-1}\simeq u^{-1}P_{0}(\Delta ^{2}/u)$. 
For the kernel defined by Eq.~(\ref{kernel}), the characteristic decay $%
\mathcal{Z}(u)$ results  
\end{subequations} 
\begin{equation} 
\mathcal{Z}(u)=\frac{1}{u+C_{1}+C_{\alpha }u^{1-\alpha }},\ \ \ \ \ \ \ 
C_{\alpha }=\frac{\beta ^{1-\alpha }\Delta ^{2\alpha }}{\gamma }. 
\label{zenodecay} 
\end{equation}%
As can be seen in Figure 3 (dotted line), besides the oscillatory behavior, 
after the transient $\gamma t\ll 1$, this function provides an excellent 
fitting of the decay dynamics.  
\begin{figure}[tbph] 
\includegraphics[height=7.4cm]{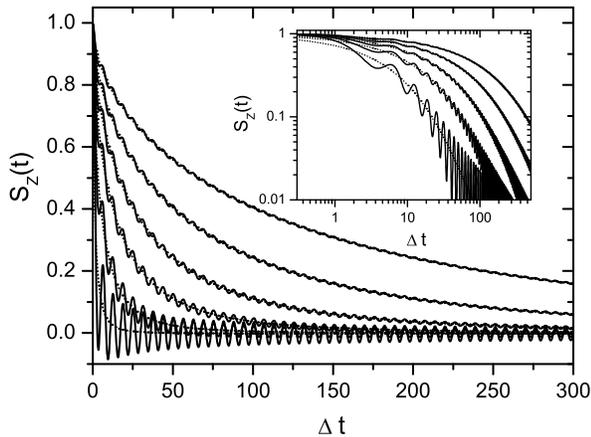} 
\caption{Average $S_{Z}(t)$ considering as initial condition the upper 
eigenstate of $\protect\sigma _{z}$. The fitting decay curves (dotted lines) 
are given by Eq.~(\protect\ref{zenodecay}). From top to bottom, the 
parameters are $\protect\gamma /\Delta =200$, $100$, $50$, $25$, $10$, and $%
2.5$. In all cases we take $\protect\alpha =1/2,$ $\protect\beta =\protect%
\gamma /2,$ and $\protect\omega _{A}=0.$ In the inset we show the same 
graphic in a log-log scale.} 
\label{Figura3} 
\end{figure} 
 
The function $\mathcal{Z}(t)$ is characterized by a reach variety of 
behaviors. First, we note that in the Markovian limit, $\beta =0$, we get an 
exponential decay with rate $C_{1}=\Delta ^{2}/\gamma $, which clearly 
diminish by increasing $\gamma $. In the non-Markovian case, in a short time 
regime, we can approximate%
\begin{equation} 
\mathcal{Z}(t)\simeq \exp \{-[C_{1}t+\frac{C_{\alpha }t^{\alpha }}{\Gamma 
(1+\alpha )}]\}, 
\end{equation}%
while in an asymptotic long time limit we get%
\begin{equation} 
\mathcal{Z}(t)\simeq \frac{(1-\alpha )}{\Gamma (\alpha )}\frac{C_{\alpha }}{%
C_{1}^{2}}\frac{1}{t^{2-\alpha }}. 
\end{equation}%
Thus, the dispersion of the random rate (measured by $\beta $) induce, at 
short times, an extra stretched exponential decay, while in the asymptotic 
regime it scales a power law behavior [$C_{1}^{2}/C_{\alpha }=\Delta 
^{2(2-\alpha )}/\gamma \beta ^{1-\alpha }$]. The characteristic rates of 
both regimes arise from a competence between the unitary and dissipative 
dynamics. We notice that by increasing the dispersion rate $\beta $, the 
characteristic rate of the stretched exponential decay is increased, while 
the rate for the power law regime is decreased. The dependence in the 
hopping frequency $\Delta $ is the inverse one. 
 
The Zeno-like effect can be \textit{qualitatively} understood in terms of 
the stochastic evolution corresponding to the effective approximation Eq.~(%
\ref{aproximada}). This stochastic process develops in the system Hilbert 
space and consists in the application at random times of the superoperator $%
\mathcal{E}=\mathcal{L}+$\textrm{I}, which in view of Eq.~(\ref{lindblad}) 
reads $\mathcal{E}[\bullet ]=\sigma _{z}\bullet \sigma _{z}$, while in the 
intermediates times the system evolves with its unitary evolution $U(t)=\exp 
[-i\Delta t\sigma _{x}/2].$ The superoperator $\mathcal{E}$ implies the 
disruptive transformations $S_{X}\rightarrow -S_{X}$, $S_{Y}\rightarrow 
-S_{Y}$, $S_{Z}\rightarrow S_{Z}$, while the unitary dynamics is equivalent 
to a rotation around the $x$-direction. In the limit of vanishing hopping 
frequency $\Delta $, the continuous applications of the superoperator $%
\mathcal{E}$ kill the $x$-$y$ components and frozen the dynamics in the 
initial condition $S_{Z}(0)$. Thus, a pure Zeno effect is recuperated. For $%
\Delta /\langle \gamma _{R}\rangle \ll 1,$ the decay dynamics is determined 
from the competence between the transformations induced by $\mathcal{E}$ and  
$U(t)$, defining the Zeno-like regime. This interpretation is exact in a 
Markovian limit and always valid for the effective master equation Eq.~(\ref%
{aproximada}). 
 
\subsection{Anomalous decay behavior from a finite set of random rates} 
 
In obtaining the previous results we have assumed an infinite set of random 
rates, Eq.~(\ref{expo}), whose effects can be approximated by the kernel 
Eq.~(\ref{kernel}). While this election guarantees the presence of an 
asymptotic power law decay, strong non-exponential behaviors can be obtained 
in an \textit{intermediate regime} by considering only a finite set, $%
1<R\leq N_{\max }$, of random rates $\{\gamma _{R},P_{R}\}$. On the other 
hand, for a finite set, the asymptotic system dynamics is always Markovian 
and characterized by the inverse rate $\langle 1/\gamma _{R}\rangle $. This 
result follows from $\lim_{u\rightarrow 0}K(u)=\langle 1/\gamma _{R}\rangle.$ 
 
In Figure~4 we show the decay dynamics induced by an environment 
characterized by a finite set of random rates $\gamma _{R}$ ($N_{\max }=7$) 
with equal weights, $P_{R}=1/N_{\max }$. Each curve follows from a 
superposition of Markovian solutions, Eq.~(\ref{Markovian}) with $\gamma 
\rightarrow \gamma _{R}$. The set of rates $\{\gamma _{R}\}$\ of each plot 
differ in a multiplicative factor, in such a way that the relation between 
the average rate $\gamma =\langle \gamma _{R}\rangle $ and the corresponding 
fluctuation rate $\beta =[\langle \gamma _{R}^{2}\rangle -\langle \gamma 
_{R}\rangle ^{2}]/\langle \gamma _{R}\rangle $\ remains constant in all 
curves. For the case $\gamma /\Delta =2.5,$ the random rates are $\gamma 
_{R}/\Delta =0.59$, $1.0$, $1.09$, $1.21$, $4.0$, $4.7$ and $4.88$.  
\begin{figure}[tbph] 
\includegraphics[height=7cm]{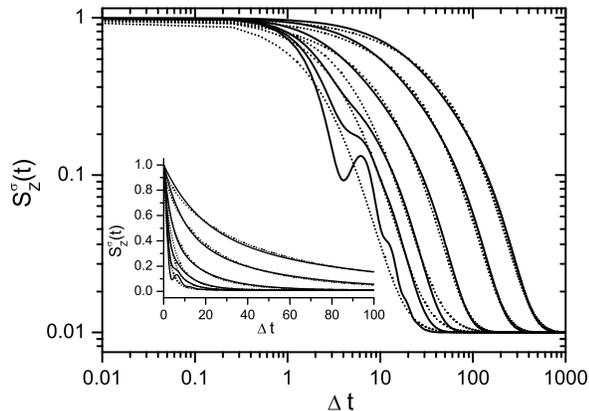} 
\caption{Average $S_{Z}^{\protect\sigma }(t)=\protect\sigma +(1-\protect%
\sigma )S_{Z}(t),$ considering as initial condition the upper eigenstate of $%
\protect\sigma _{z}$ and a finite set of random rates with equal weights. 
From top to bottom, the parameters are $\protect\gamma /\Delta =50$, $25$, $%
10$, $5$, $3.5$, and $2.5$. In all cases we take $\protect\sigma =0.01$, $%
\protect\beta /\protect\gamma =0.51,$ and $\protect\omega _{A}=0.$ For $%
\protect\gamma /\Delta \geq 10,$ the dotted lines correspond to the fitting $%
S_{Z}^{\protect\sigma }(t)\simeq \protect\sigma +(1-\protect\sigma )\exp [-(%
\protect\zeta t)^{\protect\delta }]$, while for $\protect\gamma /\Delta \leq 
10,$\ they corresponds to $S_{Z}^{\protect\sigma }(t)\simeq \protect\sigma %
+(1-\protect\sigma )[1+\protect\zeta t]^{-\protect\delta }$ (see text).} 
\label{Figura4} 
\end{figure} 
 
In order to enlighten the intermediate non-exponential regime, we have 
plotted the shifted average $S_{Z}^{\sigma }(t)=\sigma +(1-\sigma )S_{Z}(t)$%
, with $\sigma \ll 1$. In the deep Zeno-like regime [$\gamma /\Delta \gtrsim 
10$], $S_{Z}^{\sigma }(t)$ can be well approximated by an stretched 
exponential behavior $S_{Z}^{\sigma }(t)\simeq \sigma +(1-\sigma )\exp 
[-(\zeta t)^{\delta }]$, with $\delta \approx 0.7$ and $\zeta /\Delta \in 
(0.025,0.12)$. For $\gamma /\Delta \lesssim 10$, a power law fitting is more 
adequate $S_{Z}^{\sigma }(t)\simeq \sigma +(1-\sigma )[1+\zeta t]^{-\delta 
}, $ with $\delta \in (3.5,4.5)$ and $\zeta /\Delta \in (0.05,0.125)$. We 
note that a similar non-exponential fitting was found in Ref. \cite{wong} by 
considering the action of a finite bath, which can be associated with a 
glassy environment. On the other hand, the oscillatory effects in the decay 
of $S_{Z}^{\sigma }(t)$ arise from the Markovian solutions, Eq.~(\ref%
{Markovian}), corresponding to the rates satisfying $\gamma _{R}/\Delta \leq 
2$. In fact, for the Markovian solution, this condition delimits the change 
between a monotonous and an oscillatory decay behavior. Consistently, we 
notice that by increasing the average rate, the amplitude of the 
oscillations are smaller. A similar effect can be seen in Figure~3. 
 
\subsection{Mapping with other models} 
 
Our formalism relies on the applicability of the generalized Born-Markov 
approximation. Here we explore the possibility of mapping its dynamics with 
other models that also induce anomalous decay behaviors. 
 
\textit{Spin-Boson model}: The spin-boson model is defined by the total 
Hamiltonian  
\begin{equation} 
H_{T}=\frac{\hbar }{2}\{\omega _{A}\sigma _{z}+\Delta \sigma _{x}\}+\frac{%
\hbar }{2}\sigma _{z}Q_{B}+H_{B}, 
\end{equation}%
where the bath Hamiltonian $H_{B}=\sum_{j}[(p_{j}^{2}/2m_{j})+m_{j}\omega 
_{j}^{2}q_{j}^{2}]$ corresponds to a set of harmonic oscillators, and $%
Q_{B}=d\sum_{j}\kappa _{j}q_{j}$. The bath is characterized by the spectral 
density function%
\begin{equation} 
J(w)=(\pi /2)d\sum\nolimits_{j}(\kappa _{j}^{2}/m_{j}w_{j})\delta (w-w_{j}), 
\end{equation}%
and assumed to be in equilibrium at temperature $T$. As is well known, the 
reduced system dynamics can be obtained in an exact way \cite%
{leggett,grifoni,lang,hanggi,milena}. It reads  
\begin{subequations} 
\label{spin-boson} 
\begin{eqnarray} 
S_{X}(t) &=&\int_{0}^{t}d\tau \lbrack Y_{A}^{(s)}(t-\tau 
)+Y_{A}^{(a)}(t-\tau )S_{Z}(\tau )]\ \ \ \ \ \ \  \\ 
&&+Y_{B}^{(s)}(t)S_{X}(0)+Y_{B}^{(a)}(t)S_{Y}(0),  \notag \\ 
S_{Y}(t) &=&\frac{1}{\Delta }\frac{dS_{Z}(t)}{dt}, \\ 
\frac{dS_{Z}(t)}{dt} &=&\int_{0}^{t}d\tau \lbrack K_{A}^{(a)}(t-\tau 
)-K_{A}^{(s)}(t-\tau )S_{Z}(\tau )]\ \ \ \ \ \ \   \notag \\ 
&&+K_{B}^{(a)}(t)S_{X}(0)+K_{B}^{(s)}(t)S_{Y}(0)\}, 
\end{eqnarray}%
where the corresponding kernels can be written as functions of $J(w)$. On 
the other hand, it is possible to write the exact averaged evolution Eq.~(%
\ref{bloch}) in the form Eqs.~(\ref{spin-boson}). In Appendix B we present 
the kernels corresponding to each dynamics. From these expressions, it is 
simple to proof that to first order in $\Delta $, after disregarding a phase 
factor, both set of kernels can be mapped under the condition  
\end{subequations} 
\begin{equation} 
\sum_{R}P_{R}\ \exp [-\gamma _{R}t]=\exp [-Q^{\prime }(t)],  \label{map} 
\end{equation}%
where%
\begin{equation} 
Q^{\prime }(t)=\frac{d^{2}}{\hbar \pi }\int_{0}^{\infty }dw\frac{J(w)}{w^{2}}%
\coth (\frac{\hbar w}{2kT})[1-\cos (wt)], 
\end{equation}%
define the real part of the thermal\ bath correlation. We remark that the 
mapping Eq.~(\ref{map}) is only valid in a high temperature limit, condition 
consistent with the Lindblad structure Eq.~(\ref{lindblad}). 
 
In this context, from Eq.~(\ref{map}), it is possible to enlighten the 
difference between the present approach and that developed in Refs. \cite%
{tannor,ulrich}. In our approach, which relies in splitting the Hilbert 
space of the bath as a \textit{direct sum} of subspaces, $\exp [-Q^{\prime 
}(t)]$ is written as a sum of exponential functions, each one associated to 
each Markovian sub-reservoir. Instead, in Refs. \cite{tannor,ulrich}, $%
Q^{\prime }(t)$ is expressed as a sum of exponential functions. This 
representation relies in an artificial discomposing of the spectral density 
function $J(w)$ as a sum of individuals terms. Thus, the Hilbert space of 
the bath is effectively split in an \textit{external product} of subspaces, 
each one associated to a non-Markovian sub-reservoir. As in our approach, 
the system density matrix can be written in terms of a set of auxiliary 
sub-density-matrixes. Nevertheless, their evolution involves coupling among 
them all. 
 
\textit{Stochastic Hamiltonian}: Decoherence in small quantum systems is 
also modeled by introducing stochastic elements in the system evolution. 
This situation arises naturally in many physical systems \cite%
{makhlinReport,makhlin,falci,adrian}. Consistently with the spin-boson model 
we consider the stochastic Hamiltonian%
\begin{equation} 
H_{st}(t)=\frac{\hbar }{2}\{[\omega _{A}+\xi (t)]\sigma _{z}+\Delta \sigma 
_{x}\}, 
\end{equation}%
where $\xi (t)$ is a classical non-white noise term. 
 
By assuming $\langle \langle \xi (t)\rangle \rangle _{\xi }=0$, where $%
\langle \langle \cdots \rangle \rangle _{\xi }$ means an average over 
realizations of the noise, in the limit of vanishing $\Delta $ it is simple 
to solve the stochastic dynamics and obtain the average of the Pauli 
matrixes. The final evolution is the same as in Eq.~(\ref{disperso}) after 
replacing $P_{0}(t)$ with the average dephasing factor $D(t)=\langle \langle 
\exp [i\int_{0}^{t}d\tau \xi (\tau )]\rangle \rangle _{\xi }$. Thus, the 
generalized Born-Markov approximation can be mapped to the stochastic 
Hamiltonian evolution\ under the condition $D(t)=P_{0}(t)$, which explicitly 
reads  
\begin{equation} 
\sum_{R}P_{R}\ \exp [-\gamma _{R}t]=\Big\langle\Big\langle\exp 
[i\int_{0}^{t}d\tau \xi (\tau )]\Big\rangle\Big\rangle_{\xi }. 
\end{equation}%
This condition can be consistently satisfied if the dephasing factor $D(t)$ 
decays in a monotonous way. 
 
\section{Summary and Conclusions} 
 
We have presented a theoretical approach intended to describe the dynamic of 
small quantum systems interacting with a complex structured environment. Our 
formalism is based in an extension of the well known Born-Markov 
approximation, which relies in the possibility of splitting the environment 
as a direct sum of sub-reservoirs, each one being able to induce by itself a 
Markovian system dynamics. Then, we have demonstrated that the full action 
of the complex environment can be described through a random Lindblad master 
equation. The set of random rates follows from a Fermi golden rule. Thus, 
they are proportional to the characteristic coupling strength of each 
subspace multiplied by the corresponding sub-density of states evaluated in 
a characteristic frequency of the system. The associated probabilities are 
defined by the weight of each subspace in the stationary state of the bath. 
 
From a phenomenological point of view, the set of random rates and weights 
can be determined in a consistent way in function of the system decay. In 
fact, the system dynamics is characterized by a non-Markovian master 
equation that in function of the random rate set can develop strong 
non-exponential decays. 
 
As an example we worked out the dissipative dynamic of a quantum tunneling 
system in a two level approximation. We have introduced a set of random 
rates that lead to the presence of asymptotic power law decay. In the limit 
of small hopping frequency, when compared with the average rate, we have 
showed that a Zeno-like phenomenon arises, which is characterized by a 
stretched exponential and a power law decay. These behaviors follow from the 
interplay between the unitary dynamics and the entanglement-memory-effects 
induced by the reservoir. 
 
For the tunneling dynamics, we have also demonstrated that non-exponential 
decays arise even by considering a small set of random rates. Furthermore, 
we have established the conditions under which the random Lindblad evolution 
can be mapped to a spin-boson model and a stochastic Hamiltonian evolution. 
 
Finally, we want to emphasize that the present results define a new 
framework for describing anomalous quantum system dynamics, which consists 
in taking the characteristic rate of a Lindblad equation as a random 
distributed variable. We remark that this approach was not derived from an 
ensemble of identical systems whose local interactions with the environment 
can be approximated by different Markovian evolutions. In fact, the 
underlying microscopic physics can be related to a \textit{single} quantum 
system coupled to an environment with a complex structured spectral density 
function and whose dynamical influence over the system can be approximated 
by a direct sum of Markovian sub-reservoirs. Thus, our approach may be 
relevant for the description of anomalous decay processes in individual 
mesoscopic systems embedded in a condensed phase environment \cite%
{schlegel,brokmann,grigolini}. A natural example for which the generalized 
Born-Markov approximation may applies are glassy reservoirs, where the 
underlying configurational disorder produce a hierarchical distribution of 
coupling strength between the single system and the corresponding localized 
eigenstates of the reservoir \cite{wong}. 
 
\appendix 
 
\section{Exact kernels} 
 
Here we present the exact expressions for the kernels $\Gamma _{X}(u)$, $%
\Gamma _{Y}(u)$, and $\Upsilon (u)$ that define the evolution of the Pauli 
operators average, Eqs.~(\ref{bloch}). For arbitrary rates ${\{\gamma 
_{R},P_{R}\}}$, the kernels read  
\begin{subequations} 
\label{exactkernels} 
\begin{eqnarray} 
\Gamma _{X}(u) &=&D\{[u(u+C)+\Delta ^{2}](u+B)+u\omega _{A}^{2}\},\ \ \ \ \  
\\ 
\Gamma _{Y}(u) &=&D\{[u(u+B)+\Delta ^{2}](u+C)+u\omega _{A}^{2}\},\ \ \ \ \  
\\ 
\Upsilon (u) &=&D(B-C)u\omega _{A},\ \ \ \ \  
\end{eqnarray}%
where $D$ denotes the function  
\end{subequations} 
\begin{equation} 
D(u)=\frac{B(u)}{\{u[u+B(u)]+\Delta ^{2}\}[u+B(u)]+u\omega _{A}^{2}}. 
\end{equation}%
The extra functions $B$ and $C$ are defined by%
\begin{equation} 
B(u)=\frac{\left\langle G(u)\gamma _{R}\right\rangle }{\left\langle 
G(u)\right\rangle },\ \ \ \ \ C(u)=\frac{\left\langle G(u)\gamma 
_{R}^{2}\right\rangle }{\left\langle G(u)\gamma _{R}\right\rangle }, 
\end{equation}%
where we have introduced%
\begin{equation} 
G(u)=\frac{1}{[u(u+\gamma _{R})+\Delta ^{2}](u+\gamma _{R})+u\omega _{A}^{2}}%
. 
\end{equation}%
Using that the Laplace transform of $f(t)e^{\pm i\omega _{A}t}$ is given by $%
f(u\mp i\omega _{A})$, in the case $\Delta =0$ it is possible to recuperate 
the expressions of Section III-A, Eqs.~(\ref{kernelDispersivo}). On the 
other hand, taking $\omega _{A}=0$ it is straightforward to get the results 
of Section III-C, Eqs.~(\ref{kernelHoping}). 
 
In an effective approximation, Eq.~(\ref{aproximada}), the corresponding 
kernels read  
\begin{subequations} 
\begin{eqnarray} 
\Gamma _{X}(t) &=&K(t)\{(\frac{\Delta }{\varphi })^{2}+(\frac{\omega _{A}}{%
\varphi })^{2}\cos [\varphi t]\}, \\ 
\Gamma _{Y}(t) &=&K(t)\cos [\varphi t], \\ 
\Upsilon (t) &=&K(t)\frac{\omega _{A}}{\varphi }\sin [\varphi t], \\ 
\Phi _{X}(t) &=&K(t)\frac{\omega _{A}\Delta }{\varphi ^{2}}\{1-\cos [\varphi 
t]\}, \\ 
\Phi _{Y}(t) &=&K(t)\frac{\Delta }{\varphi }\sin [\varphi t], 
\end{eqnarray}%
where $\varphi =\sqrt{\omega _{A}^{2}+\Delta ^{2}}.$ 
 
The extra kernels $\Phi _{X}(t)$ and $\Phi _{Y}(t)$\ couples the derivative 
of $S_{Z}(t)$ to the averages $S_{X}(t)$ and $S_{Y}(t)$ respectively, i.e., $%
dS_{Z}(t)/dt=\Delta S_{Y}(t)-\int_{0}^{t}d\tau \{\Phi _{X}(t-\tau 
)S_{X}(\tau )+\Phi _{Y}(t-\tau )S_{Y}(\tau )\}.$ For the exact evolution, 
these kernels vanish. 
 
\section{Kernels in the spin-boson-model notation} 
 
The kernels of the spin-boson model Eqs.~(\ref{spin-boson}), in lowest order 
in $\Delta $ read \cite{milena}  
\end{subequations} 
\begin{subequations} 
\label{kernelspinboson} 
\begin{eqnarray} 
Y_{A}^{(s)}(t) &\simeq &-\Delta Y_{B}^{(s)}(t)\sin [Q^{\prime \prime }(t)], 
\\ 
K_{A}^{(a)}(t) &\simeq &\Delta ^{2}Y_{B}^{(a)}(t)\sin [Q^{\prime \prime 
}(t)], \\ 
K_{A}^{(s)}(t) &\simeq &\Delta ^{2}Y_{B}^{(s)}(t)\cos [Q^{\prime \prime 
}(t)], \\ 
Y_{A}^{(a)}(t) &\simeq &-\Delta Y_{B}^{(a)}(t)\cos [Q^{\prime \prime }(t)], 
\\ 
K_{B}^{(s)}(t) &\simeq &\Delta Y_{B}^{(s)}(t), \\ 
K_{B}^{(a)}(t) &\simeq &-\Delta Y_{B}^{(a)}(t), \\ 
Y_{B}^{(s)}(t) &\simeq &\cos [\omega _{A}t]e^{-Q^{\prime }(t)}, \\ 
Y_{B}^{(a)}(t) &\simeq &-\sin [\omega _{A}t]e^{-Q^{\prime }(t)}, 
\end{eqnarray}%
where $Q^{\prime }(t)$ and $Q^{\prime \prime }(t)$ are defined by  
\end{subequations} 
\begin{eqnarray} 
Q^{\prime }(t) &=&\frac{d^{2}}{\hbar \pi }\int_{0}^{\infty }dw\frac{J(w)}{%
w^{2}}\coth (\frac{\hbar w}{2kT})[1-\cos (wt)],\ \ \ \ \ \  \\ 
Q^{\prime \prime }(t) &=&\frac{d^{2}}{\hbar \pi }\int_{0}^{\infty }dw\frac{%
J(w)}{w^{2}}\sin (wt). 
\end{eqnarray}%
The exact evolution Eq.~(\ref{bloch}) can be written as in Eq.~(\ref%
{spin-boson}) with the definitions  
\begin{subequations} 
\label{kerneldisorder} 
\begin{equation} 
Y_{A}^{(s)}(u)=K_{A}^{(a)}(u)=0, 
\end{equation} 
\begin{eqnarray} 
Y_{A}^{(s)}(u) &=&K_{A}^{(a)}(u)=0, \\ 
K_{A}^{(s)}(u) &=&T(u)\Delta ^{2}[u+\Gamma _{X}(u)], \\ 
Y_{A}^{(a)}(u) &=&T(u)\Delta \lbrack \omega _{A}-\Upsilon (u)], \\ 
Y_{B}^{(s)}(u) &=&T(u)\Delta \lbrack u+\Gamma _{X}(u)], \\ 
K_{B}^{(a)}(u) &=&T(u)\Delta \lbrack \omega _{A}-\Upsilon (u)], \\ 
Y_{B}^{(s)}(u) &=&T(u)[u+\Gamma _{Y}(u)], \\ 
Y_{B}^{(a)}(u) &=&-T(u)[\omega _{A}-\Upsilon (u)], 
\end{eqnarray}%
where we have introduced  
\end{subequations} 
\begin{equation} 
T(u)=\frac{1}{[\omega _{A}-\Upsilon (u)]^{2}+[u+\Gamma _{X}(u)][u+\Gamma 
_{Y}(u)]}. 
\end{equation}%
The structure of these kernels is the same as those of the spin-boson model 
in the limit of vanishing $\Delta $, which implies that $\Delta $ only 
appears through the unitary evolution. In fact, in this limit we can 
approximate Eqs.~(\ref{exactkernels}) by $\Gamma _{X}(u)\simeq \Gamma 
_{Y}(u)\simeq \lbrack K(u-i\omega _{A})+K(u+i\omega _{A})]/2$ and $\Upsilon 
(u)\simeq \lbrack K(u-i\omega _{A})-K(u+i\omega _{A})]/2i$. After 
introducing these expressions in Eqs.~(\ref{kerneldisorder}), it is simple 
to get $Y_{B}^{(s)}(t)\simeq \cos [\omega _{A}t]P_{0}(t)$, and $%
Y_{B}^{(a)}(t)\simeq -\sin [\omega _{A}t]P_{0}(t)$. Then, disregarding in 
Eqs.~(\ref{kernelspinboson}) the phase contribution proportional to $%
Q^{\prime \prime }(t)$, which is valid in a high temperature limit \cite%
{wilkie}, a mapping with Eqs.~(\ref{kerneldisorder}) can be done after 
imposing the equality $P_{0}(t)=e^{-Q^{\prime }(t)}$.


\begin{thebibliography}{99}
\bibitem{footnotebath} Notice that when a bath can be split as an external 
product of Markovian sub-reservoirs, the corresponding reduced system 
dynamics is Markovian.
\bibitem{footnote0} The stochastic dynamics can be formally extended to the 
case $\mathcal{L}\neq \mathcal{E}-$I \cite{budini}\ after introducing the 
superoperator $\mathcal{E}[\rho ]=\{$I$+[e^{\kappa \mathcal{L}}-$I$]\}\rho $%
, where $\kappa $ must be intended as a control parameter. Then, the 
evolution Eq.~(\ref{aproximada}) is recuperated in the limit in which 
simultaneously $\kappa \rightarrow 0$ and the number of events by unit of 
time go to infinite, the last limit being controlled by the sprinkling 
distribution $f(t)$ \cite{CoolingBook}. This object is defined as the 
probability density for an event at time $t$, disregarding the possibility 
of extra events in $(0,t)$. Therefore, it is given by\ $f(t)=w(t)\theta 
(t)+\int_{0}^{t}w(t-\tau )f(\tau ),$ where $\theta (t)$ is the step function 
[here defined as $\theta (t)=1$ for $t>0$, and $\theta (t)=0$ for $t\leq 0$, 
which implies $f(t)|_{t=0}\equiv 0].$ In the Laplace domain it reads $%
f(u)=w(u)/[1-w(u)]$, which allows to write $K(t)=df(t)/dt$.
\bibitem{footnote1} When the random rate\ assumes only two different values $%
\gamma _{\uparrow /\downarrow }$, with probabilities $P_{\uparrow 
/\downarrow }$, the exact kernel reads $K(u)=\langle \gamma \rangle 
/[1+\beta \sigma (u)]$ where $\sigma (u)=1/[u+\eta /(\langle \gamma \rangle 
\langle \tau \rangle )]$. The rate $\beta $ is defined in terms of the 
fluctuations $\beta =[\langle \gamma _{R}^{2}\rangle -\langle \gamma 
_{R}\rangle ^{2}]/\langle \gamma _{R}\rangle .$ The extra rates are $\eta 
=P_{\uparrow }\gamma _{\downarrow }+P_{\downarrow }\gamma _{\uparrow }$, and  
$\langle \tau \rangle =P_{\uparrow }/\gamma _{\uparrow }+P_{\downarrow 
}/\gamma _{\downarrow }$.
\end{thebibliography}
\end{document}